\documentstyle[preprint,aps,eqsecnum]{revtex}

\newcommand\ignore[1]{}

\begin{document}

\draft
\title{Current and charge distributions \\of the fractional quantum Hall
liquids with edges} \author{Jun'ichi Shiraishi
and Mahito Kohmoto} \address{Institute for Solid State Physics, 
University of Tokyo, Roppongi, Minato-ku, Tokyo 106, Japan}

\maketitle
\begin{abstract}
An effective Chern-Simons theory for the quantum Hall 
states with edges is studied by treating the edge and bulk properties in a 
unified fashion. 
An exact steady-state solution is obtained for a
half-plane geometry using the Wiener-Hopf method. For a 
Hall bar with finite
width, it is proved that the charge and current distributions do not have a
diverging singularity. 
It is shown that there exists only a single mode
even for the hierarchical states,
and the mode is not localized
exponentially near the edges. Thus 
this result differs from the edge picture
in which electrons are treated as  strictly 
one dimensional chiral Luttinger liquids. 
\end{abstract}

\narrowtext

\section{Introduction}

We consider the large scale physics of the fractional quantum 
Hall (FQH) liquids with boundaries.
Our analysis is based on 
the effective Chern-Simons gauge field theory
\cite{ZHK,Zhang,SAfrica,WZR,bw,WN,WZ91,ezawa,FKT,FZ}.

In an approach initiated by Wen \cite{WTF} (see also Ref. \cite{FKT}), 
which is different from ours, a one-dimensional edge action is added to the 
original action in order to assure the gauge invariance for the 
Chern-Simons gauge field
for restricted  geometries. This has been 
followed by a number of authors
\cite{bw,Wen,LW,KFP,Haldane} who tried to explain the FQH effect based 
only on the 1d theory ({\it edge picture}) \cite{Butt,MacD}.  It has an
attractive feature, if correct, that the celebrated  Tomonaga-Luttinger
liquids are realized at the edges of FQH systems \cite{KF,FN,Moon71}. 
In those theories, however, apparently the global properties like
distributions of current, charge  and electromagnetic fields are assumed to
be insignificant. Thus one of the main consequences is that the charges and  
currents are localized near the edges. 

Recently Nagaosa and Kohmoto \cite{NK} 
examined  boundary conditions for FQH liquids and 
it was shown that  the gauge
invariance is not violated if one imposes
a physically suitable boundary
condition. Thus one  needs not to modify 
the original action as was done by Wen.
In this way,
it was possible  to study the edge and bulk 
properties on an equal footing. They
showed the fractional quantization of 
Hall conductance by considering both edge and bulk.

We begin with the composite fermion
picture of the FQH effect in Sec. II. Then the  Chern-Simons effective
field theory to describe the composite fermions and the boundary
condition are introduced in Sec. III. 

Following the line of Ref. \cite{NK} the charge and current distributions of the
FQH liquids in steady states  are discussed in Sec. IV. It is shown that
the effect of the interaction can not be treated perturbatively from 
the non-interacting situations.  The approach we are taking
for this correlated electron problem is totally non-perturvative. The
following two geometries are considered:  (i) half-plane  (subsection A) and
(ii) Hall bar (subsection B). 

\noindent(i) An exact solution  is obtained using the Wiener-Hopf
method. This method was  used, in the same
geometry,  by Thouless \cite{Thouless} to solve the
equations of MacDonald, Rice and
Brinkman \cite{MRB} for the
non-interacting integer quantum Hall liquid.

\noindent (ii) The asymptotic behavior near the edges
are obtained using the theorems of the singular integral equations 
( Hilbert
transformation).  It is shown
that there exists only a single mode even for the hierarchial 
states. {\it
This mode is  not  localized exponentially along the boundaries
and our results are
not consistent with the edge picture of the FQH effect}.

It  is also shown that
the charge distribution 
is {\it finite}  over the sample 
including the edges  
This is indispensable to obtain a well-defined theory,
since the  electron number density must be positive. 
If the charge density has singularities \cite{MRB}, one can not avoid 
negative divergence in the electron number density which leads to an  
ill-defined theory.

\section {Composite fermion picture of Fractional Hall liquids}

Filling factor is defined by
\begin{equation}
\nu = {N_e\over N_\phi} \label{filling}
\end{equation}
where $N_e$ is the number of electrons and $N_\phi$ is the number 
of flux quanta. For $\nu=1$ the first Landau level is totally filled 
and the
higher levels are all empty. Let us  see why Hall  liquids with 
inverse
filling factor equal to an odd integer might be special. Recall that 
in this
case the background magnetic field contains
an odd number of magnetic flux quanta per electron.

The first argument we cite is due to Jain.\cite{J1,J2} Imagine an adiabatic
process in which we somehow move some of the flux quanta so that $p$
units of flux are attached to each electron. For $p$ even,
the additional Dirac-Aharonov-Bohm phase associated with moving one
electron
around another is $e^{i \pi p} = 1$ and so the statistics of the electron is
unchanged. The electrons are now moving in a reduced magnetic field
$B_{\rm{eff}} = B - 2 \pi p n$ where $n$ is the number density of
electrons.
(Note that in
 our convention,  the unit of flux is $2\pi$.)
The filling factor has been increased to $\nu_{\rm{eff}}$, given by
$\nu_{\rm{eff}}^{-1} = (B - 2 \pi p n) / 2 \pi n = \nu^{-1} - p$.
For $\nu_{\rm{eff}} = m$ an integer, we have $\nu^{-1} = p + m^{-1}$
and
\begin{equation}
\nu \enskip = \enskip \frac{m}{m p + 1}\,.                       \label{11.9}
\end{equation}
Thus, fractional Hall systems with
$\nu = m/(m p +1)$ \ ($p$ even) may be adiabatically changed into an
integer Hall system with filling factor $m$, as was also emphasized by
Greiter and
Wilczek \cite{GW}. Note that this argument gives us more
than we had
hoped for. The case  we wanted to understand,
with $\nu^{-
1}=$ an  odd integer, is obtained for $m = 1$.

In this way a Hall
liquid with $\nu = m/(m p +1)$
with $m$ an integer and $p$ an even
integer, is related to the integer
Hall
liquid. We may thus want to argue that since the integer Hall liquid is
incompressible the fractional Hall liquid is also incompressible.

Another argument, historically earlier than Jain's
argument,  is due to Zhang, Hansson, and
Kivelson\cite{ZHK}. Imagine attaching all of the flux to the electrons.
Thus, each electron gets attached to it an odd number of flux quanta. By
our preceding argument, the electrons with the attached flux quanta
become bosons. We now have bosons moving in the absence of a
background magnetic field. Since bosons can condense according to Bose and
Einstein, we conclude that a Hall liquid with inverse filling factor equal to an
odd integer is energetically favored.

\section{Effective field theory}

The situation described in Sec. II above may be effectively represented
by the Chern-Simons gauge field theory coupled with external 
electromagnetic potential. The Chern-Simon term plays a role of attaching 
fluxes to
electrons. By solving the equation of motion and 
the Maxwell equation consistency, we can obtain 
the charge density, current and potential profiles.

The effective Chern-Simons gauge  Lagrangian density in the dual
representation \cite{WZDD1,mpaf,WZDD2} is 

\begin{eqnarray}
{\cal L} &=&    
{ 1 \over {4\pi} } 
\sum_{I,J} K_{IJ} \varepsilon^{\mu \nu \lambda}
a_{I \mu} \partial_{\nu} a_{J \lambda}
- \sum_I
\left(
 { 1 \over {2\pi}} A_{\mu} 
 \varepsilon^{\mu \nu \lambda}
 \partial_{\nu} a_{I \lambda}
 + { 1 \over g} f_{I \mu \nu} f^{\mu \nu}_I 
\right) ,                                                    \label{eqlag}
\end{eqnarray}
where $a_{I\mu}$ is the Cherm-Simons gauge field, the integer-valued 
symmetric matrix $K$ is written $K=I+p\,C$, where $I$ is  the $m\times m$
identity matrix and $C$ is the $m\times m$  matrix in which every element is
unity, namely, $$
C=
\left(
\begin{array}{cccc}
1&1& \cdots&1\\
1&1&        & \vdots\\
\vdots&  & \ddots&\\
1& &\cdots  & 1
\end{array}
\right).
$$
The matrix $K$ specifies the 
coupling among the Chern-Simons gauge fields and is also 
related to the filling factor  by 
$$
\nu=\sum_{IJ} (K^{-1})_{IJ}
$$ 
of FQH liquid \cite{WZk}.
The Maxwell term $ ( 1/ g ) f_{I \mu \nu} f^{\mu \nu}_I $ 
($ f_{I \mu \nu} = \partial_{\mu} a_{I \nu} - \partial_{\nu} a_{I 
\mu}$ )
in (\ref{eqlag}) is explicitly written as 
$ ( 2 / g) [c^2 f_{I xy}^2 - f_{I0x}^2 - f_{I0y}^2]$, 
where $g$ is the coupling constant and $c$ is the velocity of the 
Bogoliubov mode \cite{Zhang}. 
The vector potential $A_{\mu}$ ($\mu=0,x,y$)
of the electromagnetic field 
is coupled to the $\mu$-th component of the charge current density which is
given
by  
\begin{equation}
J^{\mu}=\sum_{I=1}^m J_{I}^{\mu},    \label{Jel}
\end{equation}
where the contribution from I-th conserved current densities is
\begin{equation}
J^{\mu}_I = { 1 \over 2\pi} \varepsilon^{\mu \nu \lambda} 
\partial_{\nu} a_{I \lambda}. \label{JI}
\end{equation}

Note that the vector potential 
for the constant external magnetic field $ B_0$
has been already taken into account in the structure of the $K$ 
matrix, and 
is not included in $A_{\mu}$. Similarly $a_{I \mu}$ and the density 
$J^0_I$ 
are measured from their average values in the following discussion.  

The Lagrangian density is integrated over a sample 
$S$. 
On the boundary $\partial S$
we impose
\begin{equation}
 \sum_{\alpha = x, y} J_I^{\alpha} n_{\alpha} |_{\partial S}= 0,\label{boundary}
\end{equation}
where $\vec n = ( n_x, n_y)$ is the unit vector normal to the 
boundary.
This boundary condition simply expresses the physical condition 
that the current can not flow through the boundary $\partial S$. 
Since it is a physical requirement, it is obviously invariant
with respect to a gauge
transformation $a_{I \mu} \to a_{I \mu} + \partial_\mu \phi_I$.
A remarkable fact is that the Chern-Simons term in Lagrangian
density (\ref{eqlag})  is also gauge invariant after integration over
$S$ with the boundary condition (\ref{boundary}).

\section{Charge and current distribution}

We consider a FQH liquid with filling 
$\nu=m/(mp+1)$ where $m$ is an integer and $p$ is 
an even integer.
To study steady-state distributions of
 charges and currents,
we need the equation of motion derived from the effective action for the
Lagrangian density (\ref{eqlag}). It reads

\begin{equation}
\sum_J K_{IJ} J_J^{\mu}
= 
\frac{1}{2\pi}\varepsilon^{\mu \nu \lambda}\partial_{\nu} 
A_{\lambda} 
+ 4\partial_{\nu} f^{\mu \nu}_I/g  \label{eqmot}
\end{equation} 
which is expressed in terms of the current density only and is gauge 
invariant.
It is explicitly written
\begin{equation}
\left[
\begin{array}{ccc}
K & - 8\pi g^{-1} \partial_t & - 8 \pi c^2 g^{-1} \partial_{y} \\ 
8\pi g^{-1} \partial_{t}& K & 8\pi c^2 g^{-1} \partial_{x} \\
- 8\pi g^{-1} \partial_y & 8 \pi g^{-1} \partial_x & K 
\end{array} \right]
\left[
\begin{array}{c}
{\bf J}^x \\
{\bf J}^y \\
{\bf J}^0 
\end{array} \right]
={ 1 \over {2 \pi}}
\left[
\begin{array}{c}
(\partial_y A_0-\partial_t A_y ) {\bf q} \\
-(\partial_x A_0-\partial_t A_x  ){\bf q} \\
 B {\bf q} 
\end{array}
\right]  , \label{eqofmotion0}
\end{equation}
where ${\bf J^{\mu}} = ^t[J_1^{\mu}, \cdot \cdot , J_m^{\mu}]$ 
and ${\bf q} = ^t[1, \cdot \cdot , 1] $ are 
vectors with $m$ components. 

Let us investigate the stational distributions of 
potential and charge in 
a Hall bar which is uniform in the $y$-direction, then one may  put 
$
\partial_t J_I^{\mu}=\partial_t A_{\mu}=0
$
in (\ref{eqofmotion0}). From the homogeneity of the system in the
$y$-direction,  impose
$
\partial_y J_I^{\mu}=\partial_y A_{\mu}=0$.
Thus we have
\begin{eqnarray}
\displaystyle
K {\bf J}^y
+
\displaystyle\frac{8\pi c^2}{g}
\partial_x {\bf J}^0 
&=&
-\displaystyle\frac{1}{2\pi}
 \partial_x A_0 {\bf q}, \label{eqm1}\\
\displaystyle\frac{8\pi }{g} \partial_x {\bf J}^y
+
K {\bf J}^0 
&=&
0. \label{eqm2}
\end{eqnarray}
Combining these, we have 
an equation for ${\bf J}^0 $:
\begin{equation}
K^2 {\bf J}^0 
-
\left(\frac{8\pi c}{g}\right)^2
\partial_x^2 {\bf J}^0 
=
\frac{4}{g} \partial_x^2 A_0 {\bf q}, \label{eqforj0}
\end{equation}
Here the potential $A_0$ is connected to
the charge current 
$J^0 (x)\equiv \sum_I J^0_I(x)$ through
\begin{equation}
A_0(x)
=
-
\xi \int_{L_1}^{L_2} dx' \ln |x-x'| J^0 (x')     \label{eqpot}
\end{equation}
where $\xi$ is a constant having the dimension of velocity
$[\frac{L}{T}]$.
Once ${\bf J}^0$ is determined,  the current ${\bf J}^y$ is obtained 
from (\ref{eqm1}) or (\ref{eqm2}).

To diagonalize these matrix equations, introduce an orthogonal matrix
$$
U=
\left(
\begin{array}{ccccc}
1/\sqrt{2} &0& \cdots &0&-1/\sqrt{2}\\
0&1/\sqrt{2}& \cdots&0&-1/\sqrt{2} \\
\vdots&  & \ddots&\vdots&\vdots\\
0&\cdots&0&1/\sqrt{2}&-1/\sqrt{2} \\
1/\sqrt{m}&1/\sqrt{m} &\cdots  & 1/\sqrt{m}  & 1/\sqrt{m}
\end{array}
\right).
$$
The matrix $C$ is diagonalized as
$$
U C U^{-1}=
\left(
\begin{array}{cccc}
0 &0& \cdots &0\\
0&0& \cdots&0 \\
\vdots&  & \ddots&\vdots\\
0&0&\cdots  & m
\end{array}
\right).
$$
Then (\ref{eqforj0}) becomes decoupled equations
\begin{eqnarray}
&&(1+mp)^2 J^0 (x)
-
\left(\frac{8\pi c}{g}\right)^2
\partial_x^2 J^0 (x)
=
\frac{4m}{g}
\partial_x^2 A_0(x), \label{eqphys}\\
&&
(J_I^0-J_m^0)
-
\left(\frac{8\pi c}{g}\right)^2
\partial_x^2 (J_I^0-J_m^0)=0
\qquad
{\rm for }\;I=1,\cdots,m-1. 
\end{eqnarray}
The densities $J_I^0-J_m^0$ ($I=1,\cdots,m-1$) 
which are orthogonal to $J^0 (x)$
and sometimes called as ``neutral modes" are
unphysical degrees of freedom, since
they do not have electro-magnetic couplings.

Therefore there is only a single mode even for the hierarchical FQH
states and it is not exponentially localized near the edge as shown
below. This result is in  sharp contrast with the claims of Wen
\cite{Wen} and Macdonald \cite{MacD} that there exist a number of
edge branches in the hierarchical FQH liquids.

By solving (\ref{eqphys}), all the long-range
behavior of the electronic density and current can be obtained.
We have two characteristic length scales

\begin{eqnarray}
&&\lambda_1=\frac{8\pi c}{g (1+mp)  },
\end{eqnarray}
and
\begin{eqnarray}
&&\lambda_2={4\pi m\xi \over g(1+mp)^2}. \label{lscale}
\end{eqnarray}
These two scales should be much larger 
than the magnetic length scale since 
our starting point is
the long-range effective theory
described by the Lagrangian density (\ref{eqlag}). 
The length scale $\lambda_1$ appears as the localization length of an edge
mode in \cite{NK}. , This edge mode, however,  does not exist when the Hall
conductance is quantized, {\it i.e.} when the longitudinal voltage drop is
zero. In what follows,  we will denote the ratio of the two scales as 

\begin{eqnarray} \eta=\left({\lambda_1 \over \lambda_2}\right)^2= \left({2c
\over \nu \gamma }\right)^2,             \label{eta} 
\end{eqnarray}
and study the effects of the parameter $\eta$.

In the Lagrangian density (\ref{eqlag}), the particle-particle 
repulsive interaction is
represented by the Maxwell term (more specifically, by the spatial part $c^2
f_{I xy}^2$). Thus if $\eta=\lambda_1/\lambda_2=0$, (\ref{eqpot}) and 
(\ref{eqphys}) represent a non-interacting case. Remarkably these
equations with  $\eta=0$ are essentially identical to those of MacDonald,
Rice and Brinkman   \cite{MRB} obtained to study the  charge and potential
profiles in the {\it integer} quantum Hall  effect. It is quite unexpected
since our method is
based on the  effective field theory and 
is completely different from theirs.

\subsection{Half-plane}

The Wiener-Hopf method is used to obtain the charge, current and potential 
profiles
of the  FQH liquids of an infinitely wide Hall bar with a single edge
located at $x=0$.
For the integer quantum Hall liquid {\it i.e.} $\eta = 0$, Thouless
\cite{Thouless} analytically solved the  equations of MacDonald {\it et al.}
by the Wiener-Hopf method in the same geometry. It is shown, however, that the
nature of the solutions are completely different for the two cases: $\eta = 0$
and $\eta \ne 0$. So the effect of the interaction can not be treated
perturbatively from  the non-interacting situations ($\eta = 0$).  The approach
we are taking for this correlated electron problem is totally non-perturbative.

In the present geometry, (\ref{eqpot})  and (\ref{eqphys})  are written
\begin{eqnarray}
&& J^0 (x)
-
\lambda_1^2
\partial_x^2 J^0 (x)
=
{\lambda_2 \over \pi\xi}
\partial_x^2 A_0(x) \qquad (0\leq x<\infty),  \label{equation2}\\
&&A_0(x)
=
-
\xi
\int_0^{\infty} dx' \ln |x-x'| J^0 (x').
\end{eqnarray}
To apply the Wiener-Hopf technique, we separate $A_0(x)$ as 
$A_0(x)=A_0^+(x)+A_0^-(x)$, where
\begin{eqnarray}
A_0^+(x)&=& 
\left\{
\begin{array}{cc}
 A_0(x) & {\rm for}\; x\geq 0,\\
 0    & {\rm for}\; x<0 ,
\end{array}
\right. 
\\
A_0^-(x)&=& 
\left\{
\begin{array}{cc}
0     & {\rm for}\; x\geq 0 ,\\
A_0(x)  & {\rm for}\; x< 0 .
\end{array}
\right.  
\end{eqnarray}
Extend (\ref{equation2}) to the region $x<0$ as
\begin{eqnarray}
 J^0 (x)
-
\lambda_1^2
\partial_x^2 J^0 (x)  
&=&
\frac{\lambda_2}{\pi\xi} \partial_x^2
( A_0^+(x) +\theta(-x) (A_0^+(0) + {A_0^+}'(0) x))  \\
&=&
\frac{\lambda_2}{\pi\xi} (\partial_x^2
 A_0^+(x) -\delta'(x) A_0^+(0) - \delta(x) {A_0^+}'(0) x)  
\quad (-\infty<x<\infty), 
\end{eqnarray}
where $\theta(x)$ is the step function:
$\theta(x)=1$ for $x>0$ and $\theta(x)=0$ for $x<0$, and
prime denotes differentiation with respect to $x$.

Introduce Fourier transforms of these functions as
\begin{eqnarray}
f(k)&=&\int_{-\infty}^{\infty} dx J^0 (x) e^{i k x},\\
g_{\pm}(x)&=&\int_{-\infty}^{\infty} dx A_0^{\pm}(x) e^{i k x}.
\end{eqnarray}
Note $f(k)$ and $g_+(k)$ converge in the upper complex $k$ plane 
and
$g_-(k)$ converges in the lower complex $k$ plane.
Regularize the Fourier transformation of the logarithm 
by multiplying a dumping factor $e^{-a|x|}$ as 
\begin{eqnarray}
&&\int_{-\infty}^{\infty} dx  e^{-a|x|}\ln |x| e^{i x k}
=
\frac{-2}{k^2+a^2}\left[
a C + \frac{1}{2} a \ln (k^2+a^2)+ k \tan^{-1}\frac{k}{a}\right],
\end{eqnarray}
where 
$\gamma=\lim_{m\rightarrow \infty} (\sum_{n=1}^{m} \frac{1}{n}-\ln m 
)$ is Euler's constant.
The r.h.s. becomes 
$-\frac{\pi}{|k|}$ if $k\neq 0$ in the limit $a\rightarrow 0$ and 
it diverges if $k=0$. By discarding this divergence, we have 
regularized equations
\begin{eqnarray}
f(k)+
\lambda_1^2\,
k^2 f(k)
&=&
\frac{\lambda_2}{\pi \xi}
(-k^2 g_+(k) -{A_0^+}'(0)+iA_0^+(0) k),\\
g_+(k)+g_-(k)
&=&
\xi
\frac{\pi  }{|k|} f(k).
\end{eqnarray}
Elliminating $g_+(k)$, we have
\begin{equation}
f(k)\left(1+ \lambda_2\, |k|
+
\lambda_1^2\,
k^2\right)
=
\frac{\lambda_2}{\pi\xi}
(k^2 g_-(k)-A_0^{+'}(0)+iA_0^+(0) k).                    \label{fk}
\end{equation}

In order to analyze (\ref{fk}) we consider factorization of the analytic
function 
\begin{equation}
1+|z|+\eta z^2=\frac{N(z)}{D(z)},   \label{factorize}
\end{equation}
where $|z|$ is defined by $|z|=z$ if ${\rm Im}(z)\geq 0$ and 
$|z|=-z$ if ${\rm Im}(z)<0$; $N(z)$ is analytic on the complex plane 
except for $\{z=i y|y\geq 0\}$ and $D(z)$ is analytic except for
$\{z=-i y|y\geq 0\}$.

Solutions for $D(z)$ and $N(z)$ are 
\begin{eqnarray}
D(z)
&=&
\eta^{-1/2} \exp \left( \frac{1}{\pi} 
\int_{0}^{\infty} i\,dy
   \frac{y \tan^{-1} \frac{y}{1-\eta y^2}}
        {iy+z}
\right),\\
N(z)
&=&
\eta^{1/2} \exp \left( \frac{-1}{\pi} 
\int_{0}^{\infty} i\,dy
   \frac{y \tan^{-1} \frac{y}{1-\eta y^2}}
        {iy-z}
\right),
\end{eqnarray}
where a branch of $\tan^{-1} y $ is chosen as $0<\tan^{-1} y<\pi$.
On the real axis $(z=k \in {\bf R})$,  $D(k)$ can be written
\begin{eqnarray}
D(k)
&=&
(1+|k|+\eta k^2)^{-1/2}
e^{i \varphi(k;\eta)}, \label{eqD}
\end{eqnarray}
where
\begin{eqnarray}
\varphi(k;\eta)
&=&
\frac{k}{\pi} 
\int_{0}^{\infty} dy
   \frac{ \tan^{-1} y-\pi/2}
        {y^2+k^2}
+
\frac{\pi}{2}(\theta(k)-1/2)\qquad {\rm for}\; \eta=0,\nonumber\\
&=&
\frac{k}{\pi} 
\int_{0}^{\infty} dy
   \frac{ \tan^{-1} \frac{y}{1-\eta y^2}-\pi}
        {y^2+k^2}
+
\pi(\theta(k)-1/2)\qquad {\rm for}\; \eta>0, 
\end{eqnarray}
Note that the phase factor $\varphi(k;\eta)$ has
a singularity at $\eta=0$.
If $\eta=0$, $\varphi(k;\eta)$ varies from $-\pi/4$ to $\pi/4$.
For $\eta>0$, however, it varies from $-\pi/2$ to $\pi/2$.

By substituting $z=\lambda_2 k$ to (\ref{factorize}) we obtain
$$
\left(1+ \lambda_2\, |k|
+
\lambda_1^2\,
k^2\right)
=
{N(\lambda_2 k) \over D(\lambda_2 k)},
$$
where $D(\lambda_2 k)$ converges 
in the upper half complex $k$-plane and 
$N(\lambda_2 k)$ converges 
in the lower half complex $k$-plane.
Rewrite (\ref{fk}) as 
$$
{f(k) \over D(\lambda_2 k)}={\frac{\lambda_2}{\pi\xi}
(k^2 g_2(k)-A'(0)+iA(0) k) \over N(\lambda_2 k)},
$$
then both sides of the equation must be well-defined. 
Therefore they are entire
functions which is a constant. Thus we have a solution 
\begin{equation}
f(k)=const.\, D(\lambda_2 k). \label{eqfD}
\end{equation}

Now we study the asymptotic behavior of $J^0 (x)$ for $\eta>0$.
The differential equation for $\varphi(k;\eta)$ is
\begin{eqnarray}
{d\varphi(k;\eta)\over dk}
&=&
{\beta_+\,\log \alpha_+ k^2 \over 2\pi(\alpha_+ k^2-1)}
+
{\beta_-\,\log \alpha_- k^2 \over 2\pi(\alpha_- k^2-1)},
\end{eqnarray}
where
\begin{eqnarray*}
\alpha_{\pm}
&=&
{2\eta^2 \over 1-2\eta\pm \sqrt{1-4\eta}}, \\
\beta_{\pm}
&=&
{\eta (\pm1 \mp 4\eta+\sqrt{1-4\eta} )\over
\sqrt{1-4\eta} (1-2\eta\pm\sqrt{1-4\eta} )  }.
\end{eqnarray*}
From this, we obtain the asymptotic forms of
$\varphi(k;\eta)$ as
\begin{eqnarray*}
\varphi(k;\eta)
&\sim&
-{1\over 2\pi}
\left(
 \beta_+ \log\alpha_+
 +
 \beta_- \log\alpha_-
\right)
k
-
{1\over \pi}k(\log k-1) +\cdots 
\qquad\qquad (k\sim 0),\\
&\sim&
{\pi \over 2}
-{1\over 2\pi}
\left(
 {\beta_+ \over \alpha_+}\log\alpha_+
 +
 {\beta_- \over \alpha_-}\log\alpha_-
\right)
{1\over k}
+
{1\over \eta\pi} {1\over k}(\log {1\over k}-1) +\cdots 
\qquad (k\sim \infty),
\end{eqnarray*}
To study the behavior of
$J^0$ on the edge,  we estimate
\begin{eqnarray}
\int_{-\infty}^{\infty} dk f(k)
&\sim&
\int_0^{c} dk (1+k+\eta k^2)^{-1/2} (c_1 k+c_2 k(\log k-1))\nonumber\\
&&+\int_{c}^{\infty} dk (1+k+\eta k^2)^{-1/2}
 (c_1' {1\over k}+c_2'{1\over  k}(\log {1\over  k}-1))\nonumber  \\
&\sim&
\int_0^{c} dk (1-{1\over 2} k) (c_1 k+c_2 k(\log k-1))\\
&&+\int_{c}^{\infty} dk \eta^{-1/2} k^{-1}
 (c_1' {1\over k}+c_2'{1\over  k}(\log {1\over  k}-1)) \nonumber \\
&<& \infty.\nonumber 
\end{eqnarray}
Here we used (\ref{eqD}), (\ref{eqfD}) and the constants $c_i$, $c_i'$ 
are obtained by combining $\alpha_{\pm}$ and $\beta_{\pm}$
suitably.
From this, it is expected that the 
charge density takes a finite value at the edge $(x=0)$.
In other words, $J(x)$ has no divergence if $\eta>0$.
Since we have the Maxwell term in the effective Lagrangian which 
partially comes from  Coulomb repulsion between electrons, 
it is a natural consequence.
Furthermore, we can obtain the fourier transform of $f(k)$, 
using approximations
$(1+k+\eta k^2)^{-1/2} \sim \eta^{-1/2} k^{-1}$ and 
$\varphi(k;\eta)\sim \pi/2 -\theta(-k)\pi -c^2/k$, as 
\begin{eqnarray}
\int_{-\infty}^{\infty} dk f(k) e^{-ikx}
&\sim&
\int_0^{\infty} dk k^{-1} \sin(kx+{c^2\over k}) \nonumber\\
&=&const. J_0 (2cx^{1/2})\qquad (x\geq 0) ,\\
&=&0 \qquad\qquad\qquad\qquad (x< 0) ,\nonumber
\end{eqnarray}
where $J_0(x)=1-{x^2\over 4}+{x^4 \over 64}-\cdots$ is the zeroth Bessel
function of the first kind.

From these and the results of Thouless \cite{Thouless}, we have
\begin{eqnarray}
J^0 (x)
&\sim& x^{-1/2}\qquad \eta=0,  \nonumber \\ 
&\sim& 1-const.\, x \qquad \eta \neq 0. \label{J0half}
\end{eqnarray}
This behavior is shown in Fig. 1.
It can be shown from (\ref{eqm2}) that

\begin{equation}
\displaystyle\frac{8\pi }{g} \partial_x J^y +(1+ mp)  J^0 = 0. 
\end {equation}
Thus the asymptotic behavior of the 
current distribution near the edge is
\begin{eqnarray}
J^y (x)
&\sim& -x^{1/2} + const. \qquad \eta=0,\nonumber \\
&\sim& -x+const. \qquad \eta \neq 0.  
\end{eqnarray}

\subsection{  Hall bar}

For a Hall bar with finite width $[-L, L]$, although time-reversal
symmetry ($T$) is broken due to the existence of the magnetic field, the system
has $TP$ symmetry where $P$ is parity. It leads to $J^0 (x)=-J^0 (-x)$ and
$A_0(x)=-A_0(-x)$.
Since the integral in (\ref{eqpot}) has a finite interval,
the Fourier transformation method used in the last subsection  for the half
plane does not work well. Thus we use  another powerful method of Hilbert
transformation. In what follows, we rescale $x$ such that the interval
[$-L,L$] becomes [$-1,1$]. Thus we should rescale the parameters as
$\lambda_i /L$. We will write these rescaled parameters as $\lambda_i$  for
simplicity.

%
%
%
%

The derivative of (\ref{eqpot}) is
represented by the Hilbert transformation as
\begin{eqnarray}
&&\partial_x A_0(x)
=
\xi \, {\rm p.v.}
\int_{-1}^{1} dy {J^0 (y) \over y-x}
=\pi\xi\,{\cal T}_x [J^0 (y) ] ,                \label{Hilbert}
\end{eqnarray}
where  ${\cal T}_x$
denotes the Hilbert transformation \begin{equation}
{\cal T}_x[f(y)]\equiv 
{\rm p.v.} \int_{-1}^{1} {dy \over \pi}{f(y) \over y-x}.
\end{equation}
Here ${\rm p.v.}\int dy$ denotes the principal value integral.
From (\ref{eqpot})  , (\ref{eqphys}) and  (\ref{Hilbert}) the equations for
currents
$J^0(x)$, and
$J^y(x)$ are
\begin{eqnarray}
J^0 (x)
-
\lambda_1^2
\partial_x^2 J^0 (x)
&=&
\lambda_2 \partial_x {\cal T}_x[J^0(y)], \label{j0eq}\\
{\lambda_1 \over c}  J^y(x)
&=&
-\lambda_1^2 \partial_x J^0(x) 
-\lambda_2 {\cal T}_x[J^0(y)]. \label{jyeq}
\end{eqnarray}

\noindent i) $\lambda_1=0$ ($\eta=0$): 
Suppose that the charge distribution $ J^0 (x)$
has singularities at the edges $x=\pm 1$, as 
it has in the half-plane Hall bar.
The singularities must be integrable, namely,
$J^0 (x)\sim (1\pm x)^{-\alpha}$ with $0< \alpha<1$.
It can be shown from Theorem II in Appendix that if
$\alpha
\neq 1/2$, the r.h.s. of (\ref{j0eq}) has singularities $\sim (1\pm
x)^{-\alpha  -1}$ which leads to a contradiction. Thus  the sigularlity
is not allowed except  $\alpha=1/2$. 
The factor $\cot (\alpha \pi)$ in
Theorem II vanishes if  $\alpha = 1/2$ and the above argument against the
existence of singularities does not hold. 
\vspace{2mm}

{\bf Proposition I.} {\it 
If a solution of (\ref{j0eq}) 
 with $\lambda_1=0$ ($\eta=0$) has singularities at 
the edges ($x=\pm 1$), 
the power of the singularity must be $-1/2$.}
\vspace{2mm}

\noindent Note that this singularity $-1/2$ coincides with the 
result for the half-plane (see (\ref{J0half}) and
\cite{Thouless}). 
\vspace{4mm}

The inverse operation of the 
Hilbert transformation (Theorem III in Appendix),
gives a systematic expansion of $J^0(x)$ for $\lambda_2>1$ as
\begin{equation}
J^0(x) = \sum_{n=0}^\infty \lambda_2^{-n} j_n(x),
\end{equation}
where 
$j(x)$'s satisfy the relation
$$
0\;
\;\;\mathop{\longleftarrow\!\!\!-\!\!\!-}^{\partial_x {\cal T}_x}\;\;
j_0(x)
\;\;\mathop{\longleftarrow\!\!\!-\!\!\!-}^{\partial_x {\cal T}_x}\;\;
j_1(x)
\;\;\mathop{\longleftarrow\!\!\!-\!\!\!-}^{\partial_x {\cal T}_x}\;\;
\cdots.
$$
At the edges $j_0(x)$ has singularities of power $-1/2$ 
and $j_n(\pm 1)=0$ for $n\geq 1$.
 Theorems I and III in Appendix
give the explicit solutions for $j_0$ and $j_1$ as
\begin{eqnarray}
j_0(x)&=&-{x\over \sqrt{1-x^2}},\\
j_1(x)&=&
-{1 \over \pi} \sqrt{1-x^2} 
\log \left({\displaystyle 1-x \over \displaystyle1+x}\right) .
\end{eqnarray}
\vspace{7mm}
Using (\ref{jyeq}), we have 
\begin{eqnarray}
{8\pi \over g(1+mp)}
J^y(x)
&=&
\lambda_2 \left(
1+{1\over \lambda_2}
\left(
{2 \over \pi} - \sqrt{1-x^2}
\right)
\right) + {\cal O}(\lambda_2^{-1}), \label{Jyeta0}\\
{1\over \pi\xi}
A_0(x)
&=&
-
\left(
{2+\pi\lambda_2 \over \pi \lambda_2} x
-
{x \over 2 \lambda_2} \sqrt{1-x^2}
-
{\sin^{-1} x \over 2 \lambda_2}
\right)+ {\cal O}(\lambda_2^{-2}). \label{a0lo}
\end{eqnarray}
The current distribution (\ref{Jyeta0}) is symmetric. It is
contrasted with the edge picture in which the currents flow in the
opposite directions at the two edges. The density, current and
potential profiles thus obtained is plotted in Fig. 2.

\noindent ii) $\lambda_1> 0$ ($\eta>0$): In this case the singularities 
of the charge distribution at the edges is 
suppressed due to the second derivative term 
in the r.h.s. of (\ref{j0eq}).
 
If $J^0 (x)$ has  power singularities  
$\sim (1\pm x)^{-\alpha}$ ($0<\alpha<1$) at the edges,
 the l.h.s. of (\ref{j0eq})
has singularities
of power $-\alpha-2$. On the other hand  
the r.h.s. of (\ref{j0eq})
has singularities
of power $-\alpha-1$. Thus power singulariteis are
forbidden. 

If $J^0 (x)$ has a logarithmic singularities
$J^0 (x)\sim \log(1-x)-\log(1+x)$ (which we will call 
as a ``simple" logarithmic singularity),
one can prove that 
(\ref{j0eq}) does not hold since

$$
\partial_x {\cal T}_x[\log(1-y)-\log(1+y)]=
-{2\over \pi}{\log(1-x)-\log(1+x) \over 1-x^2}.
$$
For another logarithmic singularities like
higher power of logarithm, for example, 
to find explicit Hilbert transformations
becomes more cumbersome.
However, it is expected that 
$$
\partial_x {\cal T}_x[{\rm logarithmic\;singularity}]\sim
{{\rm logarithmic\;singularity}\over 1-x^2},
$$
from a  power counting argument. The second derivative of this
logarithmic singularity gives a singularity with power $-2$.
Then $J^0 (x)$ can not have any logarithmic singularities which is a
contradiction. Thus we assert
 \vspace{2mm}

{\bf Proposition II.} {\it 
A solution of (\ref{j0eq}) 
 with $\lambda_1>0$ ($\eta>0$) 
has neither a power singularity
nor a (simple) logarithmic singularity at 
the edges ($x=\pm 1$).}
\vspace{2mm}

\noindent 
Since the Coulomb interaction repels particles each other, a divergent
singularity at an edge is physically unacceptable. 
This observation is consistent
with Proposition II. Thus we claim that the  charge density is finite at
the edges. Note that there is a singularity at the edge in the
noninteracting case $\eta =0$ 
(\cite{Thouless}). This is caused by the absence of
particle repulsion and disappears once an interaction is taken into account.
 
In order to obtain $J^0(x)$ we expand it in powers of $\lambda_2$ as
$$
J^0(x)=
\sum_{n=0}^{\infty} \lambda_2^n \,j_n(x).
$$
By substituting this into (\ref{j0eq}),
we obtain
\begin{equation}
j_0(x)\equiv -{\sinh (x/\lambda_1)\over \sinh (1/\lambda_1)},   
                                                            \label{j0}
\end{equation}
and 
\begin{equation}
j_1(x)-\lambda_1^2 \partial_x^2 j_1(x) 
=
\partial_x 
{\cal T}_x[j_0(y)] \label{j1eq}.
\end{equation}
The solution of the inhomogenious differential equation 
(\ref{j1eq}) is
\begin{eqnarray}
&&j_1(x) \nonumber \\
&=&
{1\over 8 \pi\lambda_1^2 \sinh(1/\lambda_1)}\times  \nonumber\\
&&
\times\left[ 
{\sinh(x/\lambda_1) \over\sinh(1/\lambda_1)} 
\left\{
(-2-\lambda_1 e^{-2/\lambda_1}+2 \lambda_1) e^{-1/\lambda_1}
{\rm Ei}\left({2\over \lambda_1}\right) \right. \right.\nonumber \\
&&
\qquad\qquad\qquad +
(2-\lambda_1 e^{2/\lambda_1}+2 \lambda_1) e^{1/\lambda_1}
{\rm Ei}\left({-2\over \lambda_1}\right)  \nonumber\\
&&
\qquad\qquad\qquad+
2(-\lambda_1 \cosh(1/\lambda_1) + 2 \sinh(1/\lambda_1) )
{\rm Chi}\left({2\over \lambda_1}\right) \nonumber\\
&&
\qquad\qquad\qquad\left.
-
2(-\lambda_1 \sinh(1/\lambda_1) + 2 \cosh(1/\lambda_1) )
{\rm Shi}\left({2\over \lambda_1}\right)
\right\}\nonumber \\
%
%
&&
+
\left\{
(-2-\lambda_1 e^{-2/\lambda_1}+ 2 \lambda_1)
\left(
e^{x/\lambda_1}
{\rm Ei}\left({1-x \over \lambda_1}\right)
-
e^{-x/\lambda_1}
{\rm Ei}\left({1+x \over \lambda_1}\right) 
\right)
\right. \\
&&
\quad\left.
+
(2-\lambda_1 e^{2/\lambda_1}+ 2 \lambda_1)
\left(
e^{-x/\lambda_1}
{\rm Ei}\left({-1+x \over \lambda_1}\right)  
-
e^{x/\lambda_1}
{\rm Ei}\left({-1-x \over \lambda_1}\right) 
\right)
\right\}  \nonumber\\
%
%
&&
+
2
\left\{
(-\lambda_1 \cosh( x/\lambda_1) +2 x \sinh(x/\lambda_1) )
\left(
{\rm Chi}\left({1-x \over \lambda_1}\right)
-
{\rm Chi}\left({1+x \over \lambda_1}\right)\right)
\right. \nonumber\\
&&
\quad\left.\left.
+
(-\lambda_1 \sinh( x/\lambda_1) +2 x \cosh (x/\lambda_1) )
\left(
{\rm Shi}\left({1-x \over \lambda_1}\right)
+
{\rm Shi}\left({1+x \over \lambda_1}\right)
\right)
\right\} \right],                                         \label{j1}
\end{eqnarray}
where ${\rm Ei}(x)$, ${\rm Shi}(x)$ and ${\rm Chi}(x)$ 
are the exponential, hyperbolic sine and hyperbolic cosine
integrals, respectively, which are explicitly written
\begin{eqnarray}
{\rm Ei}(x)  &=&- {\rm p.v.} \int_{-x}^{\infty} { e^{-t} \over t} dt,\\
{\rm Shi}(x) &=&\int_0^x { \sinh t \over t} dt,\\
{\rm Chi}(x) &=&\gamma + \ln x + \int_0^x { \cosh t - 1\over t} dt,
\end{eqnarray}
and we have used the Hilbert transformation
\begin{eqnarray}
&&{\cal T}_x [-{\sinh (y/\lambda_1)\over \sinh (1/\lambda_1)}]\nonumber\\
&=&
-
{\sinh(x/\lambda_1)   \over \pi \sinh(1/\lambda_1)}
 \left(
{\rm Chi}\left({1-x \over \lambda_1}\right) 
-
{\rm Chi}\left({1+x \over \lambda_1} \right)
\right) \\
&&
-
{\cosh(x/\lambda_1)   \over \pi \sinh(1/\lambda_1)}
\left(
{\rm Shi}\left({1-x \over \lambda_1} \right)
+
{\rm Shi}\left({1+x \over \lambda_1} \right)
\right)\nonumber.
\end{eqnarray}
Using (\ref{eqpot}) and (\ref{jyeq}), we have
\begin{eqnarray}
{8\pi\over g(1+mp)} J^y(x)
&=&
-\lambda_1^2 \partial_x \left(j_0(x)+\lambda_2 j_1(x)\right)-
\lambda_2 {\cal T}_x [j_0(y)]+ {\cal O}(\lambda_2^2), \label{etajy}\\
{1\over \xi} A_0(x)
&=&
-{\rm p.v.}\int_{-1}^1 dx' \log|x'-x|
\left(j_0(x')+\lambda_2 j_1(x')\right) + {\cal O}(\lambda_2^2).
\end{eqnarray}
Note that the current density $J^y$ in (\ref{etajy}) has no divergence
at the  edges and it is again
symmetric. It is contrasted with the edge picture in which the
currents flow in the opposite directions at the two edges. 
The lowest order approximation of $A_0(x)$ is obtained by using
\begin{eqnarray}
&&-{\rm p.v.}\int_{-1}^1 dy \log |y-x| j_0(y) \nonumber\\
&=&
{\lambda_1 \over \sinh (1/\lambda_1)}\times\nonumber\\
&&
\times 
\left\{
-\cosh(x/\lambda_1) 
\left(
{\rm Chi}\left({1-x \over \lambda_1}\right) 
-
{\rm Chi}\left({1+x \over \lambda_1} \right)
\right) \right. \nonumber\\
&&
-
\sinh(x/\lambda_1) 
\left(
{\rm Shi}\left({1-x \over \lambda_1}\right) 
-
{\rm Shi}\left({1+x \over \lambda_1} \right)
\right)  \nonumber\\
&&
\left.
+\cosh(1/\lambda_1) 
\log \left( {1-x \over 1+x}\right)  \right\}. \label{a00}
\end{eqnarray}

The electron density is plotted
in Fig. 3 (a) in the lowest order approximation ($\lambda_2 = 0$)
(\ref{j0}) and (b)
with the correction (\ref{j1}). The current distribution is plotted
in Fig. 4 (a) in the lowest order approximation ($\lambda_2 = 0$) and
(b) with the correction (\ref{etajy}). 

Note that in the lowest order 
approximation, both the electron density and the current is localized
near the edges exponentially with the localization length $\lambda_1$
This behavior, however, completely disappears again once the
corrections are taken into account  as seen from Fig. 3 (b) and Fig.
4 (b) where the bulk currents do not vanish.

The potential is plotted in Fig. 5. 
Note that in the lowest order approximation, 
$A_0(x)$ has a maximum
and a minimum  near the edges in contrast to the non-interacting case $\eta
=0$ (\ref{a0lo}).  These maximum and minimum of $A_0(x)$
survive even if the next leading contribution is 
taken into account as seen from Fig. 5 (b).

\begin{acknowledgements}
It is pleasure to thank Y. Avishai for very important help. We also thank 
T. Eguchi and  N. Nagaosa for  useful discussions.
\end{acknowledgements}

\appendix
\section{Theorems of the singular integral equations}

The following Theorems I, II and III \cite{rTri}
are used in the main text.

{\bf Theorem I.} {\it 
\begin{equation}
{\cal T}_x \left[ \left( {1-y \over 1+y}\right)^\alpha \right]
=
\cot (\alpha \pi)
\left( {1-x \over 1+x}\right)^\alpha
-
{1 \over \sin (\alpha \pi)}
\end{equation}
for $-1<\alpha<1$.} 
\\

{\bf Theorem II.} {\it
Let $f(x)$ be an $L_p$-function ($p>1$) which in a small 
neighborhood
$(-1, -1+\delta)$ ($\delta>0$) of the point $x=-1$ 
can be written in the form
$$
f(x)=A (1+x)^{-\alpha} + g(x)\quad (0\leq \alpha<1),
$$
where $A$ is a constant, $g(x)$ vanishes at $x=-1$ and
satisfies (uniformly) a Lipschitz 
condition of positive order $\epsilon$, i.e.
$$
|g(x)-g(x_0)|< K |x-x_0|^\epsilon.
$$
Then the Hilbert transform of $f(x)$ has the asymptotic 
representation
$$
{\cal T}_x[f(y)]=
A \cot (\alpha\pi) (1+x)^{-\alpha}+ O(1)\quad (x\rightarrow -1),
$$
if  $0<\alpha<1$, and the asymptotic representation
$$
{\cal T}_x[f(y)]=
-{A \over \pi} 
\log (1+x)+ O(1)\quad (x\rightarrow
-1),
$$
if $\alpha=0$.

If the point $x=-1$ is replaced by the point $x=+1$,
then all remains
the same except that $\cot (\alpha\pi)$ is changed 
into $-\cot (\alpha\pi)$ and
$-\log (1+x)$ is changed into $+\log (1-x)$.}
\\

{\bf Theorem III.} {\it If a given function $f(x)$ 
belongs to the class $L_{4/3+\epsilon}$ for sufficiently small $\epsilon>0$,
the equation 
\begin{equation}
f(x)={\cal T}_x[\phi(y)],
\end{equation}
has the solution
\begin{equation}
\phi(x)={C \over \sqrt{1-x^2}}-{\rm p.v.}\int_{-1}^1 {dy\over \pi}
\sqrt{1-y^2 \over 1-x^2}
{f(y) \over y-x} ,
\end{equation}
where $C$ is an arbitrary constant.}
\\

\end{document}